%% file: aries2.tex

\input balticw
\input psfig.sty
\def\ts{\thinspace}
\font\rmn=cmr8
\year 2002
\VOLUME {~11}
\PAGES {231--247}
\pageno=231

\TITLE={PHOTOMETRIC INVESTIGATION OF THE \hfil\break
MBM 12 MOLECULAR CLOUD AREA IN ARIES.\hfil\break II. CLOUD DISTANCE}

\AUTHOR={V. Strai\v zys, K. \v Cernis, A. Kazlauskas and V. Laugalys}

\AUTHORHEAD={V. Strai\v zys, K. \v Cernis, A. Kazlauskas, V. Laugalys}
\ARTHEAD={MBM 12 molecular cloud area in Aries. II. Cloud distance}

\ADDRESS={Institute of Theoretical Physics and Astronomy,
Go\v stauto 12, \hfil\break Vilnius 2600, Lithuania;
straizys@itpa.lt, cernis@itpa.lt, algisk@itpa.lt,\hfil\break
vygandas@itpa.lt}

\SUBMITTED={May 25, 2002}

\SUMMARY={Photoelectric magnitudes and color indices in the {\itl
Vilnius} seven-color system for 152 stars are used to investigate the
interstellar extinction in the area of the Aries molecular cloud MBM 12,
coinciding with the L1454 and L1457 dust clouds.  Spectral types,
absolute magnitudes, color excesses, interstellar extinctions and
distances of the stars are determined.  The plot of interstellar
extinction {\itl A}$_V$ versus distance shows that the dust cloud is
situated at a distance of 325 pc, at 180 pc from the Galactic
plane, and its true diameter is about 11 pc.  The interstellar
extinction law in the area is found to be normal, typical for the
diffuse dust.  Ten peculiar or unresolved binary stars and some heavily
reddened stars are detected.  }

\KEYWORDS={stars:  fundamental parameters, classification -- ISM:  dust,
extinction, clouds -- ISM:  individual objects:  MBM 12, L1454, L1457}

\printheader

\section{1. INTRODUCTION}

A high galactic latitude molecular cloud MBM 12 (Magnani, Blitz \& Mundy
1985) in Aries at one time was considered as the nearest to the Sun
star-forming region.  The molecular cloud partly coincides with two dust
clouds L1454 and L1457 (Lynds 1962).  In the same area, Lynds has
distinguished two smaller clouds, L1453 and L1458, however their
identification seems to be ambiguous.  Also, the identification of the
Lynds and MBM clouds, given by Magnani et al.  (1985), seems to be not
sufficiently correct.
\vskip0.5mm

The first estimate of the distance of the Aries dark clouds, done by
Duerr \& Craine (1982), was based on star counts on photographs of the
area in the $V$ and $I$ passbands. They found evidence for two clouds at
the distances of 200--300 pc and 500--800 pc.
\vskip0.5mm

Next estimates of the distance of MBM 12 (Hobbs, Blitz \& Magnani 1986;
Hobbs, Blitz, Penprase, Magnani \& Welty 1988) were based on the
presence or absence of the interstellar Na\ts I lines in the spectra and
the spectroscopic distances for two stars:  HD 18404 and HD 18519/20
($\epsilon$ Ari AB).  The first of them was found to be at a
spectroscopic distance of 60 pc, with no traces of the interstellar
Na\ts I. The second star (a visual binary) exhibited a strong Na I absorption,
and its distance was accepted to be 70 pc.
\vskip0.5mm

Later on, Hearty et al. (2000a,b) have directed attention that the
{\it Hipparcos} parallaxes place the star HD 18404 at much closer
distance -- 32 pc, while HD 18519/20 was found to be at $\sim$90 pc.
Thus, the limits of the MBM 12 cloud distance  were considerably
widened. Additional stars closer than 100 pc were looked for the
Na\ts I absorption but none was found.
\vskip0.5mm

The distance of the MBM 12 cloud was reconsidered by Luhman (2001) by
analyzing the data of one of the corner stars -- $\epsilon$ Ari AB,
which has previously provided the upper limit of the cloud distance.
This binary star, consisting of A2 IV and A3 IV components, is
completely unreddened, i.e., it must be in front of the dust cloud.  The
presence of interstellar Na lines in its spectrum was explained by the
neutral gas wall associated with the Local Bubble.  Luhman concluded
that the star $\epsilon$ Ari AB gives only a lower limit for the
distance to MBM 12.  After plotting the extinction values $A_V$ for 12
stars in the immediate vicinity of MBM 12 against the {\it Hipparcos}
distances, the author has found only a gradual increase of $A_V$ with
distance up to 250 pc, with no sharp rise of extinction at the supposed
distance of the cloud.  The conclusion has been done that the cloud is
farther away than 200--250 pc. By comparing the magnitudes of the
probable foreground and background M-type dwarfs, Luhman has estimated a
cloud distance of $\sim$275 pc. This value is close to distance of the
first cloud of Duerr \& Craine (1982).
\vskip0.5mm

For the determination of more exact distance to the cloud, we need
reliable distances and extinctions for stars at larger distances.  This
information can be obtained either from MK classification of stars
combined with a two-color photometry or from multicolor photometry alone
in a photometric system which is able to classify stars in two
dimensions and give the interstellar reddenings of the stars.  Such a
system, consisting of seven medium-width passbands placed in the optimum
positions for the classification of stars of all spectral types, was
developed in Vilnius in the sixties of the 20th century.  The system is
described in detail in the Strai\v zys (1977, 1992) monographs.
\vskip0.5mm


\section{2. OBSERVATIONS, REDUCTIONS AND CLASSIFICATION OF STARS}

152 Aries stars were measured in the seven passbands of the {\it
Vilnius} system in 2000--2001 with the two telescopes:  the 156 cm
telescope of the Mol\.etai Observatory in Lithuania and the 150 cm
telescope of the University of Arizona at Mt. Lemmon, Arizona.  The area
was limited by the following 2000.0 coordinates:  RA = 2$^{\rm
h}$51$^{\rm m}$ -- 3$^{\rm h}$01$^{\rm m}$, DEC = +18$\degr$30$\arcmin$
-- +21$\degr$30$\arcmin$.  The catalog of magnitudes and color indices
and the identification chart for the measured stars are given in
Kazlauskas, \v Cernis, Strai\v zys \& Laugalys (2002, Paper I).
\vskip0.5mm

 For the determination of spectral classes
and absolute magnitudes of stars two independent methods were used.
\vskip0.5mm

(1) The $\sigma Q$-method of matching 14 different interstellar
redden\-ing-free $Q$-parameters of a program star to those of about 8400
standard stars of various spectral and luminosity classes, metallicities
and peculiarity types (Strai\v zys \& Kazlauskas 1993).
The reddening-free $Q$-parameters are defined by the equation:
$$
Q_{1234} = (m_1-m_2) - (E_{12}/E_{34})(m_3-m_4), \eqno(1)
$$
and
$$
E_{k,\ell} = (m_k-m_{\ell})_{\rm {reddened}} - (m_k-m_{\ell})_{\rm
{intrinsic}}. \eqno(2)
$$
where $m$ are the magnitudes in four (sometimes three) passbands,
$m_1-m_2$ and $m_3-m_4$ are the two color indices and $E_{12}$ and
$E_{34}$ are the corresponding color excesses.  In the medium-band {\it
Vilnius} system the ratios of color excesses depend slightly on spectral
and luminosity classes, and this dependence is taken into account.  In
calculating the $Q$-parameters, we used the color excess ratios
$E_{12}/E_{34}$ corresponding to the normal interstellar extinction law
(see Strai\v zys 1992).  The extinction law, i.e. the dependence of the
extinction on the wavelength, has not been investigated in the Aries
dark clouds.  However, studies of the extinction law in the adjacent
\hbox{areas} in Perseus
and Taurus give arguments in favor of the normal law in Aries, at least
in the directions where spatial dust density is low (for a discussion
see Strai\v zys et al. 2001a,b). Lower in this section, we give an
estimate of the ratio $E_{V-K}/E_{B-V}$ which also seems to be close to
normal.
The matching of $Q$-parameters leads to a selection of some standard
stars which have a set of $Q$s most similar to those of the program
star. The match quality is characterized by
$$
\sigma Q = \pm\sqrt{{\sum_{n}^{} \Delta Q_i^2}\over n}, \eqno (3)
$$
where $\Delta Q$ are differences of corresponding $Q$-parameters of the
program star and the standard, $n$ is a number of the compared
$Q$-parameters (in our case, $n$ = 14).  If the $\sigma Q$ value is
sufficiently small (i.e., the $Q$ differences between the program and
the standard star are small), the spectral and luminosity classes of the
closest star may be prescribed to the program star.  For photometry of
Population I stars measured with the 1\% accuracy, $\sigma Q$ is usually
of the order of $\pm(0.01-0.02)$ mag.  In most cases, for the program
star we have accepted the average spectral and luminosity classes of the
best fitted standard stars (from two to five).
\vskip0.5mm

(2) Interstellar reddening-free $Q,Q$ diagrams calibrated in terms of MK
spectral classes and absolute magnitudes $M_V$ by Strai\v{z}ys et al.
(1982).  In the present study we applied the method (2) mostly for
G8--K5 giants and subgiants for which the accuracy of determination of
absolute magnitudes by the calibrated $Q_{UPY}$, $Q_{XZS}$ and
$Q_{XZS}$,$Q_{XYZ}$ diagrams is much better than by the method (1).  The
same method was used for some stars with rare spectral types for which
no analog stars were found among the 8400 standards.  \vskip0.5mm

Both methods give the accuracy of spectral class $\pm 1$ decimal
spectral subclass.  The errors of absolute magnitudes are within $\pm
0.3$ and $\pm 0.5$ mag.  \vskip0.5mm

The color excesses $E_{Y-V}$ of stars were calculated as the differences
between the observed $Y$--$V$ and the intrinsic color indices
\hbox{($Y$--$V$)$_0$} for various spectral and luminosity classes, taken
from
Strai\v zys (1992, Tables 66--69 and 73).  The distances of stars were
calculated by the equation:
$$
5 \log r = V - M_V + 5 - A_V, \eqno (4)
$$
where $A_V=R_{YV}E_{Y-V}$.  For the majority of stars the absolute
magnitudes were taken from Strai\v zys (1992) tabulation according to
the spectral and luminosity classes, with a correction of --0.1 mag,
bringing the $M_V$ scale to the new distance modulus of the Hyades
($V$--$M_V$ = 3.3, Perryman et al. 1998).
\vskip0.5mm

The $V$ magnitudes of the stars and the results of photometric
quantification (spectral types and absolute magnitudes $M_V$) are given
in Table 1. The table also contains color excesses $E_{Y-V}$,
extinctions $A_V$, distances $r$ and the classification accuracy $\sigma
Q$.  The values of distances at \hbox{$r>200$ pc} are rounded to the
nearest number multiple of 10.  In the column headed ``Other sp. types",
the table contains spectral types of the stars collected from the
literature:  the data are taken from the HD, AGK 3 and PPM catalogs.
The last column shows that the $\sigma Q$ values for the overwhelming
majority of stars are between 0.01 and 0.02 mag, i.e., their
classification accuracy is sufficiently good.  The Q,Q mark in this
column means that the star was classified by the $Q,Q$ diagrams, not by
the $\sigma Q$ method.  \vskip0.5mm

For the estimation of the ratio $R_{YV}$ in the area, we have used the
$V-K$ color indices measured in the 2MASS survey (Skrutskie et al.
1997). We have found 25 stars with $E_{Y-V}\ge 0.20$ for which $K$
magnitudes are available.  Their spectral range is from A5\ts V to K5\ts
III.  For these stars we calculated color indices $V-K$, taking $V$ from
Table 2, and color excesses $E_{V-K}$, taking the intrinsic $(V-K)_0$
from Strai\v zys (1992, Tables 22--24).  The least square solution gives
the equation (after rejection of the two reddest K5\ts III stars):
$$
E_{V-K}/E_{Y-V} = 3.229 + 0.962 (Y-V)_0 \pm 0.240 \eqno(5)
$$
with a correlation coefficient of 0.65.  This equation is not valid for
stars with ($Y$--$V$)$_0 > 0.9$, i.e. cooler than K3\ts III.  For the
transformation of the ratio $E_{V-K}/E_{Y-V}$ to $E_{V-K}/E_{B-V}$ we
need the ratio of color excesses in the {\it Vilnius} and $BV$ systems,
$E_{Y-V}/E_{B-V}$.  Due to the band-width effect, this ratio slightly
depends on spectral type of a star.  For A5--G5 stars the ratio is
$\sim$0.80 and for G8--K2 III stars it is $\sim$0.85.  Equation (5) and
the ratio $E_{Y-V}/E_{B-V}$ lead to $E_{V-K}/E_{B-V} \approx 2.8$ for
A5--G5 stars.  This value is very close to that given by the normal
interstellar extinction law (Strai\v zys 1992). According to Cardelli,
Clayton \& Mathis (1988, 1989), the form of the extinction law in the
visible and the infrared ranges, defining the ratio $R_{BV}$, is well
correlated with its form in the ultraviolet. Thus, we may accept that in
the Aries cloud the extinction law is normal, i.e. it is typical for the
diffuse dust. The same conclusion has been done by Andersson \& Wannier
(1995) by measuring the wavelength of maximum polarization for three
stars in the area (our numbers 15, 61 and 100).
\vskip0.5mm

For the normal interstellar extinction law, the ratio $R_{YV} =
A_V/E_{Y-V} = 4.16$, with a very small dependence on spectral type of
the star (Strai\v zys et al. 1996).  We used this value of $R_{YV}$ to
calculate the extinctions $A_V$ and distances $r$ given in Table 1. The
expected errors for single stars are:  $\pm$0.03 mag for $E_{Y-V}$,
$\pm$0.1 mag for $A_V$ and $\pm$15--25\% for distance (Strai\v zys et
al. 2001a).
\vskip0.5mm

\pageinsert
\Table{1}{ Results of photometric quantification, determination of color
excesses, extinctions, and distances of the stars.}
\vskip-0.3cm
$$\vbox{\tabskip 33pt minus 33pt\rmn\baselineskip=11pt
\halign to \hsize {
        \hfill # &  # & # \hfil & # \hfil &
        \hfill # \hfil & \hfill # & \hfil # \hfil &
        \hfill # & \hfil #  & \hfil # \cr
\noalign{\smallskip\hrule\smallskip}
 \hfill No. \hfill & \hfil GSC~~~~~ & Photom. \hfil &  Other &
\hfil {\itl V}~~~ & \hfill {\itl M$_V$}
\hfill & \hfill ~~{\itl E$_{Y-V}$} \hfill & \hfill {\itl A$_V$} \hfill &
\hfill {\itl r} \hfill & \hfill $\sigma${\itl Q} \hfill \cr
& & sp. type \hfil &  sp. types \hfil & mag & mag & mag & mag &
\hfil pc \hfil & \hfil mag \hfil \cr
\noalign{\medskip\hrule\medskip}
  1.*\hskip3pt & 1227:0327 &  F9 IV    &  F8   &   6.62 &  +2.8 & 0.03 & 0.12 &   55 & 0.01  \cr
  2.         & 1227:0966 &  G5 V     &  G0   &   8.97 &  +5.0 & 0.01 & 0.04 &   61 & 0.01  \cr
  3.         & 1227:0013 &  F5 IV    &  F5   &  10.62 &  +2.5 & 0.10 & 0.42 &  350 & 0.01  \cr
  4.         & 1227:0163 &  G9 III   &  K0   &   9.34 &  +0.9 & 0.13 & 0.54 &  380 & Q,Q  \cr
  5.*\hskip3pt & 1227:0222 &  K5 II-III &  K5   &   8.36 & --1.0 & 0.16 & 0.66 &  550 & Q,Q   \cr
  6.         & 1227:0280 &  F5 V     &       &  11.46 &  +3.5 & 0.07 & 0.29 &  340 & 0.01  \cr
  7.         & 1230:0673 &  G7 III-IV   &       &  11.85 &  +1.6 & 0.17 & 0.71 & 810 & Q,Q  \cr
  8.         & 1227:0305 &  G9 IV-V  &       &  11.11 &  +4.3 & 0.08 & 0.33 &  198 & Q,Q   \cr
  9.         & 1230:0729 &  M0 III   &       &  11.09 & --0.6\rlap{:} & 0.25 & 1.02 & 1360\rlap{:} & Q,Q   \cr
 10.         & 1230:0942 &  K0 III   &       &  12.35 &  +0.9 & 0.13 & 0.54 & 1520 & Q,Q  \cr
 11.*\hskip3.2pt & 1227:0649 &  F8 I-II? &  F    &   9.21 &       & &      &      &       \cr
 12.         & 1227:0081 &  K1 III   &  K0   &   9.03 &  +0.7 & 0.11 & 0.46 &  370 & 0.02  \cr
 13.         & 1227:0136 &  F8 V     &       &  12.04 &  +4.0 & 0.12 & 0.50 &  320 & 0.01  \cr
 14.*\hskip3.2pt & 1230:0782 &  G5 IV-V: &       &  11.10 &  +4.0 & 0.12\rlap{:} & 0.50\rlap{:} &  210\rlap{:} & 0.02  \cr
 15.         &  1227:0617 &   K0 III  &   G5, K0 &  8.83 &   0.0 &  0.17 &  0.71 &   420 &  Q,Q \cr
 16.         &  1227:0449 &   A5 m    &       &   10.69 &   +1.8\rlap{:} & 0.20\rlap{:} & 0.83\rlap{:} &  410\rlap{:}  & Q,Q  \cr
 17.         &  1230:0814 &   G8.5 IV   &       &   11.51 &   +2.4 &  0.16 &  0.67 &   490 &  Q,Q \cr
 18.         &  1227:0550 &   K0 III  &       &   11.77 &   --0.3 &  0.24 &  1.00 &  1640 &  Q,Q \cr
 19.*\hskip3.2pt &  1230:0724 &   G7 III  &   G5, K0 &  8.24 &   +0.8 &  0.13 &  0.54 &   240 &  0.01 \cr
 20.         &  1227:0469 &   G9.5 III &       &   11.34 &  --0.1 &  0.17 &  0.71 &  1400 &  Q,Q \cr
 21.*\hskip3.2pt &  1230:0966 &   A5 V    &       &   12.01 &   +1.8 &  0.30 &  1.25 &   620 &  0.01 \cr
 22.         &  1227:0585 &   G8 III  &   G5  &    9.71 &   +1.5 &  0.15 &  0.61 &   330 &  Q,Q \cr
 23.         &  1227:0457 &   F5 IV   &       &   10.92 &   +2.5 &  0.08 &  0.33 &   410 &  0.01 \cr
 24.         &  1227:0412 &   K0 III-IV &     &   12.15 &   +1.8 &  0.12 &  0.50 &   930 &  Q,Q  \cr
 25.         &  1227:0045 &   K1 III  &       &   12.39 &   +0.5 &  0.35 &  1.46 &  1220 &  Q,Q  \cr
 26.         &  1227:0046 &   F6 V    &       &   12.05 &   +3.6 &  0.24 &  1.00 &   310 &  0.01 \cr
 27.         &  1227:0790 &   F5 V    &       &   10.81 &   +3.5 &  0.06 &  0.25 &   260 &  0.01 \cr
 28.*\hskip3.2pt &  1227:0725 &   M4.5 III &       &   10.96 &       &      &      &      &       \cr
 29.         &  1227:0231 &   K0.5 V  &       &   12.18 &   +6.0 &  0.07 &  0.29 &   150 &  0.02 \cr
 30.*\hskip3.2pt &  1227:0218 &   F3 IV   &   A5  &    9.05 &   +2.4 &  0.05 &  0.21 &   194 &  0.01 \cr
 31.         &  1227:0428 &   A2 V    &       &   11.75 &   +1.2 &  0.14 &  0.58 &   990 &  0.01 \cr
 32.         &  1227:0131 &   K3 V    &       &   11.74 &   +6.6 &  0.04 &  0.17 &    98 &  0.02 \cr
 33.         &  1230:0767 &   G2 V    &       &   10.15 &   +4.6 &  0.05 &  0.21 &   117 &  0.01 \cr
 34.         &  1227:0073 &   K1 V    &       &   12.23 &   +6.1 &  0.05 &  0.21 &   153 &  0.02 \cr
 35.         &  1227:0460 &   G2 V    &       &   11.53 &   +4.6 &  0.06 &  0.25 &   220 &  0.01 \cr
\noalign{\medskip\hrule\medskip}
}}$$
\endinsert

\pageinsert
\line{\rml\hfil Table~1 (continued)}
\vskip-0.3cm
$$\vbox{\tabskip 33pt minus 33pt\rmn\baselineskip=11pt
\halign to \hsize {
        \hfill # &  # & # \hfil & # \hfil &
        \hfill # \hfil & \hfill # & \hfil # \hfil &
        \hfill # & \hfil #  & \hfil # \cr
\noalign{\smallskip\hrule\smallskip}
 \hfill No. \hfill & \hfil GSC~~~~~ & Photom. \hfil &  Other &
\hfil {\itl V}~~~ & \hfill {\itl M$_V$}
\hfill & \hfill ~~{\itl E$_{Y-V}$} \hfill & \hfill {\itl A$_V$} \hfill &
\hfill {\itl r} \hfill & \hfill $\sigma${\itl Q} \hfill \cr
& & sp. type \hfil &  sp. types \hfil & mag & mag & mag & mag &
\hfil pc \hfil & \hfil mag \hfil \cr
\noalign{\medskip\hrule\medskip}
 36.         &  1227:0554 &   G5 III  &       &   11.83 &   +0.9 &  0.15 &  0.62 &  1150 &  0.01 \cr
 37.*\hskip3.2pt &  1230:1048 &   K1 III &   K0  &    7.10 &   0.0 &  0.09 &  0.37 &  220 &  Q,Q \cr
 38.         &  1230:0999 & F5/8 II-III & G0  &    9.79 &       &      &      &      &       \cr
 39.         &  1230:0630 &   G5 IV   &       &   11.73 &   +3.4 &  0.19 &  0.79 &   320 &  Q,Q \cr
 40.         &  1230:1004 &   F0 V    &       &   12.33 &   +2.7 &  0.38 &  1.58 &   410 &  0.02 \cr
 41.         &  1227:0564 &   M0/1 III  &       &   10.01 &  &   &  &  & \cr
 42.         &  1230:0428 &   F7 IV-V &       &   11.10 &   +3.4\rlap{:} & 0.14 &  0.58 &   260\rlap{:} & 0.01 \cr
 43.*\hskip3.2pt &  1230:0816 &   K1.5 III&   K0  &    6.95 &   +0.2 &  0.10 &  0.42 &   184 &  Q,Q \cr
 44.*\hskip3.2pt &  1227:0916 &   F0 IV   &   A3  &    7.00 &   +2.1 &  0.01 &  0.04 &    94 &  0.01 \cr
 45.         &  1230:1055 &   F8 V    &       &   10.88 &   +4.0 &  0.04 &  0.16 &   220 &  0.01 \cr
 46.*\hskip3.2pt &  1230:0821 &   G2 V    &   F8  &    8.64 &   +4.6 &  0.02 &  0.08 &    62 &  0.02 \cr
 47.*\hskip3.2pt &  1230:0600 &   F4 V    &   F2  &    8.81 &   +3.3 &  0.04 &  0.16 &   117 &  0.02 \cr
 48.         &  1227:1316 &   F5 V    &       &   10.41 &   +3.5 &  0.03 &  0.12 &   230 &  0.01 \cr
 49.         &  1227:0447 &   K5 III  &       &   10.36 &  --0.7\rlap{:} & 0.36 &  1.50 &   820\rlap{:} & Q,Q  \cr
 50.         &  1227:0684 &   G0 V    &       &   10.43 &   +4.3 &  0.01 &  0.04 &   165 &  0.01 \cr
 51.         &  1230:0643 &   F9 V    &       &   10.89 &   +4.2 &  0.06 &  0.25 &   194 &  0.01 \cr
 52.         &  1227:0362 &   F8 V    &       &    9.93 &   +4.0 &  0.05 &  0.21 &   139 &  0.01 \cr
 53.         &  1230:0893 &   G5 II  &       &   10.93 &  --2.0 &  0.25 &  1.04 &  2390 &  Q,Q \cr
 54.         &  1227:0297 &   K0 II-III &     &   11.64 &   --1.4 &  0.65 &  2.70 &  1170 &  Q,Q  \cr
 55.*\hskip3.2pt &  1230:1002 &   K--M V  &  K6e   &   12.21 &        &       &       &       &       \cr
 56.*\hskip3.2pt &  1230:0416 &   G5 V    &       &   11.11 &   +5.0\rlap{:} & 0.07 &  0.29 &   146\rlap{:} & 0.03 \cr
 57.*\hskip3.2pt &  1227:1100 &   G9 V    &  K0   &    9.75 &   +5.8 &  0.02 &  0.08 &    59 &  0.02 \cr
 58.         &  1230:0854 &   F6 V    &   G0  &    9.79 &   +3.6 &  0.09 &  0.37 &   146 &  0.01 \cr
 59.*\hskip3.2pt &  1227:0558 &   K3/5 III? &      &   10.73 &       &      &      &      &       \cr
 60.         &  1227:0087 &   G0 V    &       &   11.84 &   +4.3 &  0.05 &  0.21 &   290 &  0.01 \cr
 61.*\hskip3.2pt &  1227:0883 &   F2   &   F   &    8.97 &   &  &  &  &  \cr
 62.         &  1227:0190 &   G2 IV-V   &       &   11.95 &   +3.4 &  0.20 &  0.83 &   350 &  0.01 \cr
 63.         &  1230:0797 &   K5 III  &       &   12.20 &   --0.5 & 0.31 &  1.29 &  1910 & Q,Q \cr
 64.         &  1227:1115 &   K2 III  &       &   11.13 &   +0.4 &  0.26 &  1.08 &   850 &  Q,Q \cr
 65.         &  1227:0777 &   F8 V    &   F5  &    9.84 &   +4.0 &  0.01 &  0.04 &   144 &  0.01 \cr
 66.         &  1230:0458 &   F5 V    &       &   11.78 &   +3.5 &  0.31 &  1.29 &   250 &  0.01 \cr
 67.*\hskip3.2pt &  1227:1401 &   F6 IV   &   F5, F6\ts V  &    5.58 &   +2.6 &  0.00 &  0.00 &    39 &  0.01 \cr
 68.*\hskip3.2pt &  1230:0784 &   F8     &       &   11.49 &   &  &  &   &  \cr
 69.*\hskip3.2pt &  1227:1362 &   K0.5 IV-V? &    &   11.61 &       &      &      &      &       \cr
 70.         &  1230:0771 &   G9 III-IV &     &   11.63 &   +1.3 &  0.24 &  1.00 &   730 &  Q,Q \cr
\noalign{\medskip\hrule\medskip}
}}$$
\endinsert

\pageinsert
\line{\rml\hfil Table~1 (continued)}
\vskip-0.3cm
$$\vbox{\tabskip 33pt minus 33pt\rmn\baselineskip=11pt
\halign to \hsize {
        \hfill # &  # & # \hfil & # \hfil &
        \hfill # \hfil & \hfill # & \hfil # \hfil &
        \hfill # & \hfil #  & \hfil # \cr
\noalign{\smallskip\hrule\smallskip}
 \hfill No. \hfill & \hfil GSC~~~~~ & Photom. \hfil &  Other &
\hfil {\itl V}~~~ & \hfill {\itl M$_V$}
\hfill & \hfill ~~{\itl E$_{Y-V}$} \hfill & \hfill {\itl A$_V$} \hfill &
\hfill {\itl r} \hfill & \hfill $\sigma${\itl Q} \hfill \cr
& & sp. type \hfil &  sp. types \hfil & mag & mag & mag & mag &
\hfil pc \hfil & \hfil mag \hfil \cr
\noalign{\medskip\hrule\medskip}
 71.         &  1227:0818 &   G8 V    &       &   10.94 &   +5.5 &  0.05 &  0.21 &   111 &  0.02 \cr
 72.*\hskip3.2pt &  1230:0302 &   F0 V    &   A5  &    8.76 &   +2.7 &  0.04 &  0.16 &   151 &  0.01 \cr
 73.         &  1227:0157 &   G9 IV   &       &   12.28 &   +2.4 &  0.23 &  0.96 &   610 &  Q,Q \cr
 74.*\hskip3.2pt &  1230:0286 &   G6-K1 III-IV? & &   13.46 &       &      &      &      &       \cr
 75.         &  1227:0917 &   F5 V    &       &   10.88 &   +3.5 &  0.12 &  0.50 &   240 &  0.01 \cr
 76.         &  1230:0446 &   G8 II   &       &   12.13 &   --2.0 &  0.55 &  2.29 &  2340 &  Q,Q  \cr
 77.         &  1230:0993 &   K0.5 III &       &   12.16 &  --0.1 &  0.29 &  1.21 &  1620 &  Q,Q \cr
 78.         &  1227:0036 &   F2 V    &       &   11.45 &   +3.0 &  0.16 &  0.66 &   360 &  0.01 \cr
 79.         &  1230:0404 &   K0 III  &       &   13.28 &   --0.6 &  0.34 &  1.41 &  3110 &  Q,Q  \cr
 80.         &  1227:0205 &   K1 III  &       &   10.12 &   +0.7 &  0.31 &  1.29 &   420 &  Q,Q \cr
 81.*\hskip3.2pt &  1230:0905 &   F8 II-III &     &   11.69 &       &      &      &      &       \cr
 82.         &  1230:0907 &   G0 V    &   G0  &   10.68 &   +4.3 &  0.07 &  0.29 &   165 &  0.01 \cr
 83.         &  1230:0910 &   G2 V    &       &   12.19 &   +4.6 &  0.06 &  0.25 &   290 &  0.01 \cr
 84.*\hskip3.2pt &  1227:0620 &   F2/5 IV-V &   F5  &    9.81 &  &  &  &  &  \cr
 85.         &  1227:0437 &   K2 III  &       &   12.20 &  --0.3 &  0.15 &  0.62 &  2380 &  Q,Q \cr
 86.*\hskip3.2pt &  1227:0769 &   G-K?    &       &   12.16 &         &       &       &      &       \cr
 87.         &  1230:0510 &   G3 V    &       &   11.55 &   +4.7 &  0.08 &  0.33 &   200 &  0.01 \cr
 88.         &  1230:0665 &   G4 V    &   G5  &    9.92 &   +4.9 &  0.03 &  0.12 &    95 &  0.01 \cr
 89.         &  1227:0115 &   K0 III  &       &   11.29 &   +0.7 &  0.16 &  0.66 &   970 &  Q,Q \cr
 90.         &  1230:0888 &   F6 III  &       &   11.80 &   +2.0 &  0.36 &  1.50 &   460 &  Q,Q  \cr
 91.         &  1230:0783 &   F6 V    &       &   10.31 &   +3.6 &  0.06 &  0.25 &   196 &  0.01 \cr
 92.         &  1230:0664 &   G8 V    &       &   12.17 &   +5.5 &  0.09 &  0.37 &   181 &  0.02 \cr
 93.         &  1227:0481 &   G0 IV-V &       &   11.36 &   +3.4 &  0.21 &  0.87 &   260 &  0.01 \cr
 94.         &  1230:0616 &   G0 V    &       &   10.63 &   +4.3 &  0.08 &  0.33 &   158 &  0.01 \cr
 95.         &  1230:0552 &   F9 V sd? &       &   11.40 &   +4.2 &  0.05 &  0.21 &   250 &  0.02 \cr
 96.*\hskip3.2pt &  1230:0400 &   F8 V    &       &   11.37 &  &  &  &   &   \cr
 97.*\hskip3.2pt &  1230:1425 &   F5 V    &   F0, F5\ts IV  &    5.73 &   +3.5 &  0.00 &  0.00 &    28 &  0.01 \cr
 98.         &  1227:0363 &   G9 III  &       &   11.55 &   +0.7 &  0.16 &  0.66 &  1090 &  Q,Q \cr
 99.*\hskip3.2pt &  1230:0850 &   G?      &       &   11.78 &       &      &      &      &       \cr
100.         &  1230:0432 &   F6 V    &       &    9.47 &   +3.6 &  0.09 &  0.37 &   126 &  0.01 \cr
101.         &  1227:0003 &   K1 V    &       &   12.21 &   +6.1 &  0.05 &  0.21 &   152 &  0.02 \cr
102.         &  1230:0332 &   G8 V    &       &   13.02 &   +5.5 &  0.08 &  0.33 &   270 &  0.01 \cr
103.         &  1227:1330 &   K2 V    &       &   10.53 &   +6.4 &  0.03 &  0.12 &    63 &  0.03 \cr
104.*\hskip3.2pt &  1230:0744 &   K2 V? & K4e  &   12.06 &       &      &      &      &       \cr
105.         &  1227:0870 &   K0 II-III &      &   11.34 &   --0.9 &  0.41 &  1.71 &  1280 &  Q,Q  \cr
\noalign{\medskip\hrule\medskip}
}}$$
\endinsert

\pageinsert
\line{\rml\hfil Table~1 (continued)}
\vskip-0.3cm
$$\vbox{\tabskip 33pt minus 33pt\rmn\baselineskip=11pt
\halign to \hsize {
        \hfill # &  # & # \hfil & # \hfil &
        \hfill # \hfil & \hfill # & \hfil # \hfil &
        \hfill # & \hfil #  & \hfil # \cr
\noalign{\smallskip\hrule\smallskip}
 \hfill No. \hfill & \hfil GSC~~~~~ & Photom. \hfil &  Other &
\hfil {\itl V}~~~ & \hfill {\itl M$_V$}
\hfill & \hfill ~~{\itl E$_{Y-V}$} \hfill & \hfill {\itl A$_V$} \hfill &
\hfill {\itl r} \hfill & \hfill $\sigma${\itl Q} \hfill \cr
& & sp. type \hfil &  sp. types \hfil & mag & mag & mag & mag &
\hfil pc \hfil & \hfil mag \hfil \cr
\noalign{\medskip\hrule\medskip}
106.         &  1227:0986 &   K0 III  &       &   10.80 &   +1.3 &  0.18 &  0.75 &   560 &  Q,Q \cr
107.*\hskip3.2pt &  1227:0210 &   K0 II-III &     &   12.05 &   --1.7 &  0.85 &  3.54 &  1100 &  Q,Q  \cr
108.         &  1227:0342 &   G9 III  &       &   10.94 &   +0.2 &  0.27 &  1.12 &  840 &  Q,Q \cr
109.         &  1230:0713 &   G6 IV   &   G5  &   10.15 &   +3.0 &  0.10 &  0.42 &   220 &  0.01 \cr
110.*\hskip3.2pt &  1230:0520 &   A-F     &       &   12.16 &       &      &      &      &       \cr
111.*\hskip3.2pt &  1227:0771  &   G2 V    &   G0  &    8.77 &   +4.6 &   0.00 &  0.00 &    68 &  0.01 \cr
112.         &  1227:0860 &   F6 V    &       &   12.11 &   +3.6 &  0.11 &  0.46 &   410 &  0.01 \cr
113.         &  1230:0862 &   K1 V    &       &   12.00 &   +6.1 &  0.08 &  0.33 &   130 &  0.02 \cr
114.         &  1230:0511 &   G3 V    &       &   11.40 &   +4.7 &  0.07 &  0.29 &   191 &  0.01 \cr
115.         &  1230:0550 &   G9 V    &   G5  &    9.74 &   +5.8 &  0.01 &  0.04 &    60 &  0.03 \cr
116.*\hskip3.2pt &  1230:0970 &   M1 III  &   M0  &    8.83 &   --0.8 &  0.47 &  1.96 &   340 &  Q,Q  \cr
117.*\hskip3.2pt &  1230:0240 &   A5   &   A3  &    6.70 &  &  & &  & \cr
118.*\hskip3.2pt &  1227:0642 &   G8 III  &   K0  &    8.24 &   +0.4 &  0.16 &  0.67 &   270 &  Q,Q \cr
119.         &  1230:0723 &   F0 IV   &       &   11.24 &   +2.1 &  0.41 &  1.70 &   310 &  0.02 \cr
120.         &  1227:0116 &   K4.5 II-III &   &    9.25 &   --1.7 &  0.14 &  0.58 &  1180 &  Q,Q  \cr
121.         &  1227:0160 &   F6 V    &       &   10.96 &   +3.6 &  0.06 &  0.25 &   260 &  0.01 \cr
122.         &  1227:0891 &   F8 V:   &       &   10.98 &   +4.0 &  0.04 &  0.16 &   230 &  Q,Q  \cr
123.*\hskip3.2pt &  1227:0027 &   F2 V    &   F0  &    7.32 &   +3.0 &  0.00 &  0.00 &    73 &  0.01 \cr
124.         &  1227:0494 &   F0 IV-V &       &   10.91 &   +2.4 &  0.12 &  0.50 &   400 &  Q,Q  \cr
125.         &  1230:0576 &   G3 V    &       &   10.94 &   +4.7 &  0.12 &  0.50 &   140 &  0.02 \cr
126.         &  1227:0820 &   F0 V    &       &   10.57 &   +2.7 &  0.19 &  0.79 &   260 &  0.02 \cr
127.         &  1230:0853 &   A7 V    &   A2  &    9.32 &   +2.2 &  0.09 &  0.37 &   220 &  0.02 \cr
128.         &  1230:1044 &   G2 V    &       &   12.06 &   +4.6 &  0.08 &  0.33 &   270 &  0.01 \cr
129.         &  1230:0987 &   G9 III  &       &   11.25 &  --0.4 &  0.33 &  1.38 &   1130 &  Q,Q \cr
130.         &  1227:0744 &   F9 V    &       &   12.16 &   +4.2 &  0.12 &  0.50 &   310 &  0.01 \cr
131.*\hskip3.2pt &  1230:0772 &   F7 V  &   &  10.07 &  &  &  &  &  \cr
132.         &  1227:0232 &   G8 IV   &       &   11.36 &   +2.9 &  0.16 &  0.66 &   360 &  Q,Q \cr
133.         &  1230:0655 &   G3 V    &       &   12.11 &   +4.7 &  0.11 &  0.46 &   240 &  0.01 \cr
134.         &  1230:0722 &   G7 V    &       &   10.55 &   +5.4 &  0.00 &  0.00 &   107 &  0.02 \cr
135.         &  1230:0852 &   G9.5 III &       &   12.46 &   +0.7 &  0.23 &  0.96 &  1440 & Q,Q \cr
136.         &  1230:0742 &   F5 V    &       &   10.74 &   +3.5 &  0.09 &  0.37 &   240 &  0.02 \cr
137.         &  1230:0867 &   K0 III  &       &   12.05 &   +0.4\rlap{:} &  0.25 &  1.04 &  1320\rlap{:} & Q,Q \cr
138.*\hskip3.2pt &  1227:1178 &   B9.5 V  &   A0  &    6.73 &  &  &  &  &  \cr
139.         &  1230:0679 &   G7 III  &       &   10.08 &   +1.4 &  0.25 &  1.04 &   340 &  Q,Q \cr
140.         &  1230:0735 &   M2 III  &       &   12.06 &   --1.0 &  0.31 &  1.29 &  2260 &  0.02 \cr
\noalign{\medskip\hrule\medskip}
}}$$
\endinsert

\pageinsert
\line{\rml\hfil Table~1 (continued)}
\vskip-0.3cm
$$\vbox{\tabskip 33pt minus 33pt\rmn\baselineskip=11truept
\halign to \hsize {
        \hfill # &  # & # \hfil & # \hfil &
        \hfill # \hfil & \hfill # & \hfil # \hfil &
        \hfill # & \hfil #  & \hfil # \cr
\noalign{\smallskip\hrule\smallskip}
 \hfill No. \hfill & \hfil GSC~~~~~ & Photom. \hfil &  Other &
\hfil {\itl V}~~~ & \hfill {\itl M$_V$}
\hfill & \hfill ~~{\itl E$_{Y-V}$} \hfill & \hfill {\itl A$_V$} \hfill &
\hfill {\itl r} \hfill & \hfill $\sigma${\itl Q} \hfill \cr
& & sp. type \hfil &  sp. types \hfil & mag & mag & mag & mag &
\hfil pc \hfil & \hfil mag \hfil \cr
\noalign{\medskip\hrule\medskip}
141.         &  1230:1017 &   G0 V    &       &   11.97 &   +4.3 &  0.09 &  0.37 &   290 &  0.01 \cr
142.         &  1230:1010 &   F2.5 IV &   A5  &    9.90 &   +2.3 &  0.08 &  0.33 &   280 &  0.02 \cr
143.*\hskip3.2pt &  1230:0434 &   M0/2 III &       &   11.48 &       &      &      &      &   \cr
144.         &  1230:0696 &   G8.5 IV   &   K0  &    9.98 & +2.7 &  0.09 &  0.37 &  240 &  Q,Q \cr
145.         &  1230:0694 &   G7 IV   &       &   12.07 &   +3.1 &  0.18 &  0.75 &   440 &  0.01 \cr
146.         &  1230:0786 &   F4 V    &       &   12.16 &   +3.4 &  0.25 &  1.04 &   350 &  0.01 \cr
147.         &  1230:0997 &   K3 V    &       &   11.58 &   +6.6 &  0.02 &  0.08 &    95 &  0.02 \cr
148.         &  1231:1650 &   F1 IV   &   F0  &    9.17 &   +2.2 &  0.04 &  0.16 &   230 &  0.02 \cr
149.         &  1231:1230 &   F6 V    &       &   12.14 &   +3.6 &  0.30 &  1.25 &   290 &  0.01 \cr
150.         &  1228:1069 &   F5 IV   &       &   10.42 &   +2.5 &  0.08 &  0.33 &   330 &  0.01 \cr
151.         &  1228:1624 &   F0 V    &   F0  &    9.49 &   +2.7 &  0.08 &  0.33 &   196 &  0.01 \cr
152.*\hskip3.2pt  &  1228:1051 &   F0 V    &   F2  &    9.00 &   +2.7 &  0.10 &  0.42 &   150 &  0.02 \cr
\noalign{\medskip\hrule\medskip}
}}$$
\vskip2mm
\noindent NOTES:
\vskip5mm

\tabfont{

\noindent\hskip3pt ~~1. HD 17659 (F8) = HIC 13269.

\noindent\hskip3pt ~~5. HD 17768 (K5) = HIC 13337.

\noindent~11. HD 17834 (F)

\noindent~14. Classification uncertain: photometry of low accuracy.

\noindent~19. HD 17870 (G5) = HIC 13418

\noindent~21.  According to B.-G. Andersson (personal communication) its
spectral

type is B7\ts V-III.  Further verification is necessary.

\noindent~28. Uncertain absolute magnitude.

\noindent~30. HIC 13496.

\noindent~37. HD 18019 (K0) = HIC 13532.

\noindent~43. HD 18066 (K0) = HIC 13571.

\noindent~44. HD 18091 (A3) = HIC 13579.

\noindent~46. HD 18106 (F8) = HIC 13589, standard star.

\noindent~47. HD 18090 (F2).


\noindent~55.  Classification uncertain. T Tauri type star (Hearty et al.
2000a,

 Jayawardhana et al. 2001).

\noindent~56.  Low accuracy of classification due to larger {\itl U--V}
and {\itl P--V} errors.

\noindent~57. HIC 13631.

\noindent~59.  Classification uncertain:  low accuracy observations of
{\itl U--V}, {\itl P--V} and

{\itl X--V}.

\noindent~61. HD 18190 (F). Binary WDS 02558+1909, $\rho$=0.5$\arcsec$,
$\Delta${\itl m}=1.1 mag.

\noindent~67. HD 18256 (F5) = HIC 13702, $\rho$ Ari.
}
\endinsert

\topinsert
\tabfont{

\noindent\hskip3pt ~68. Classification uncertain. Binary?

\noindent\hskip3pt ~69. Classification uncertain. Carbon-rich?

\noindent\hskip3pt ~72. HD 18283  (A5) = HIC 13723.

\noindent\hskip3pt ~74.  Classification uncertain:  low accuracy of
{\itl U--V}, {\itl P--V} and {\itl X--V}.

\noindent\hskip3pt ~81. Classification uncertain. Binary?

\noindent\hskip3pt ~84. Classification uncertain.

\noindent\hskip3pt ~86. Classification uncertain: low accuracy of {\itl
U--V}.

\noindent\hskip3pt ~96.  Binary WDS 02580+2124, $\rho$=4.6$\arcsec$,
$\Delta${\itl m}=0.1 mag.

\noindent\hskip3pt ~97. HD 18404 (F0) = HIC 13834.

\noindent\hskip3pt ~99. Classification uncertain. Binary?

\noindent 104. Classification uncertain. T Tauri type star (Hearty et al.
2000a,

Jayawardhana et al. 2001).

\noindent 107. The most reddened star in our sample.

\noindent 110. Classification impossible. Binary?

\noindent 111. HIC 13855.

\noindent 116. SAO 75669. The star has been analyzed by Bhatt et al.
(1994).

 Polarization 3.7\% (Bhatt \& Jain 1992).

\noindent 117.  HD 18484 (A3) = HIC 13892.  Binary WDS 02589+2137,
$\rho$=0.5$\arcsec$,

$\Delta${\itl Hp}=0.2 mag.

\noindent 118. HD 18485 (K0) = HIC 13893.

\noindent 123. HD 18508 (F0) = HIC 13913.

\noindent 131. Binary WDS 02597+2013, $\rho$=5.3$\arcsec$,
$\Delta${\itl m}=0.3 mag.

\noindent 138.  HD 18654 (A0) = HIC 14021, 50 Ari.  Binary WDS 03005+1800,

$\rho$=2.2$\arcsec$, $\Delta${\itl Hp}=3.0 mag.

\noindent 143. Classification uncertain. Binary?

\noindent 152. HIC 14201. }
\endinsert

18 stars from our list have trigonometric parallaxes measured by the
{\it Hipparcos} orbiting observatory.  Their HIC numbers are given in
the Notes to Table 1. Among them are two binary stars, not classified
photometrically.  The standard deviation of the distances for the 10
stars, which are closer than 200 pc, is only $\pm$7 pc.  Both distance
scales do not show any systematic differences.

\section{3. INTERSTELLAR EXTINCTION AND CONCLUSIONS}

The dependence of extinction vs. distance for the stars in the
investigated area is shown in Figure 1. We have not tried to divide the
area into smaller regions of different run of extinction vs. distance,
since the number of stars is insufficient for a detailed study.  Some
stars with low classification accuracy can be identified in Table 1 by
the absence of absolute magnitudes, color excesses and distances and the
presence of notes.  These stars were not used in the extinction study.
Probably, most of them are unresolved binaries.  They need further
individual investigation by spectral analysis, radial velocity
measurements, spectral energy distribution modeling, etc.
\vskip0.5mm

\WFigure{1}{\psfig{figure=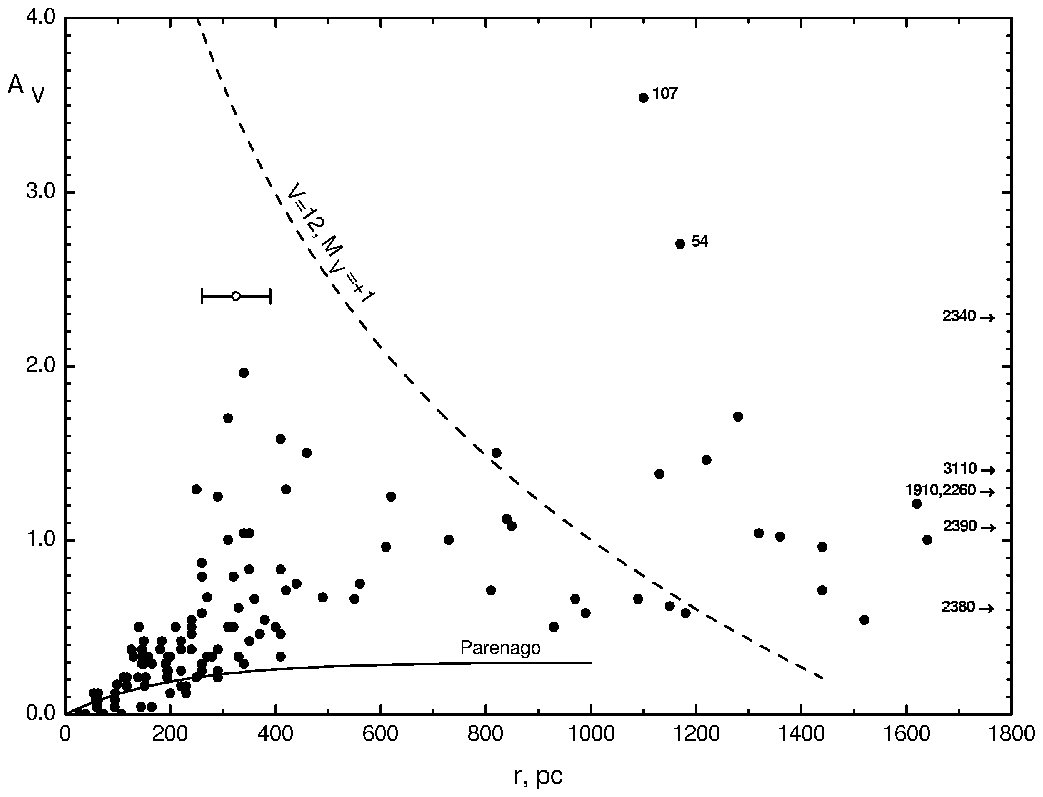,width=12truecm,angle=0,clip=}}
{~The extinction {\itl A}$_V$ plotted as a function of the distance
{\itl r} in parsecs.  The solid line shows the extinction run according
to the Parenago formula for the --34$\degr$ Galactic latitude.  The
broken line is the limiting curve for stars of {\itl V}$_{\rm {lim}}$ =
12 mag and {\itl M}$_V$ = +1 mag.  Above this curve a strong selection
effect is present.  For more details see the text.  The circle with the
error bar is shown at the estimated distance of the dust cloud.}

The following conclusions may be drawn from Figure 1.
\vskip0.5mm

(1) The non-zero extinction starts at $\sim$60 pc and it increases
gradually up to a distance of 250 pc, reaching a mean value of 0.3 mag.
This value is close to the extinction created by the general Galactic
dust layer predicted by the exponential Parenago formula with the
coefficients:  $A_0$ = 1.5 mag/kpc and $\beta$ = 0.11 kpc (Parenago
1945, Sharov 1963), shown in Figure 1.
\vskip0.5mm

(2) At distances $>$250 pc many reddened stars appear with
extinctions between 0.5 and 2.0 mag.  They are situated near the edges
of the dark clouds.
\vskip0.5mm

(3) The three most reddened stars, found in the area, are 107 (K0
II-III, $A_V$ = 3.54 mag), 54 (K0 II-III, $A_V$ = 2.70 mag) and 76 (G8
II, $A_V$ = 2.29 mag).  All the three are very distant objects ($r>1$
kpc), being seen through a lane of lower extinction near the +20$\degr$
declination circle, separating the L1454 and L1457 dust clouds.  The
star 76 is seen only at 5$\arcmin$ distance from the well-known T Tauri
type star WY Ari. Other four T Tauri type stars are in the same lane.
\vskip0.5mm

(4) There are 21 stars with distances $>$1 kpc in our sample, including
the three stars described above. All these stars are seen
through more transparent cloud sections, exhibiting extinction
values between 0.5 and 3.5 mag.  At these distances, the stars with
larger extinctions are too faint to be detected with the present
limiting magnitude.
\vskip0.5mm

(5) Figure 1 also shows the curve, corresponding to a limiting apparent
magnitude of $V$ = 12 mag and the absolute magnitude $M_V$ = +1.0.  This
value of $M_V$ corresponds to K-type giants, which are most numerous
absolutely brightest stars in the area.  Above the limiting curve, a
strong selection effect takes place:  K giants, as well as absolutely
fainter stars, with larger extinctions are too faint to be included in
the present program.  Above the curve only some stars of fainter
apparent brightness or absolutely brighter than $M_V$ = +1.0 are present.
\vskip0.5mm

We confirm Luhman's (2001) conclusion that the Aries cloud cannot be
closer than 200--250 pc.  In this distance range there is only a gradual
increase of extinction related to the general Galactic dust layer.  Only
a small hump of extinction from 0.2 to 0.4 mag at 140--160 pc may be
suspected.  It may be related to an extension of the Taurus dark cloud
complex to the Aries area.  The first considerably reddened stars start
to appear at 250 pc. However, this does not mean that the dust cloud
begins there.  The reddened stars at this distance may appear due to
accidental negative distance determination error:  at 310 pc the rms
error is about $\pm$60 pc.  Consequently, the dust cloud may be
situated at 310 pc distance or even farther.
\vskip0.5mm

Another estimate of the dark cloud distance is the distribution of stars
with low reddening.  The stars near the Parenago exponential curve
disappear at $\sim$410 pc, what means that the dust cloud rises
their positions in Figure 1 upward at a distance of 410 -- 70 = 340 pc,
here 70 pc is the distance rms error at 340 pc. This distance is not
very different from the value obtained from the considerably reddened
stars discussed in the previous paragraph.
\vskip0.5mm

Consequently, our study gives evidence that the dark cloud distance is
between 310 and 340 pc.  As a preliminary value, the distance 325
pc may be accepted.  Since the angular diameter of the dust cloud is
about $2\degr$, at this distance the true diameter is $\sim$11 pc.
Probably, its radial extent is of the same order. The distance from the
Galactic plane at $b=-34\degr$ is $\sim$180 pc. These values should be
considered as preliminary, since it is not excluded, that in future some
reddened stars at the distances closer to the Sun will be found.
\vskip0.5mm

Zimmermann \& Ungerechts (1990), Pound et al.  (1990) and
Moriarty-Schieven et al.  (1997) from the observed equivalent widths of
radio lines of $^{12}$CO and $^{13}$CO molecules have determined the
mass of the Aries cloud, which ranges between 30 and 200 $M_{\odot}$,
assuming a cloud distance of 65 pc.  If we place the cloud 5 times
farther, it becomes much more massive, 750--5000 $M_{\odot}$.  Even this
new mass is still too small in comparison with the virial mass estimated
by the same authors, assuming virial equilibrium between the
gravitational potential and the kinetic energy.  This means that the
mass of the whole cloud is not sufficient to bind it gravitationally.
A considerable fragmentation is found in the cloud with different clumps
having different velocities.
\vskip0.5mm

In our earlier paper (Strai\v zys et al. 2001b), we have plotted a polar
diagram ``distance versus galactic latitude'' for the Galactic
anticenter direction with positions of the investigated dust clouds in
Taurus, Perseus and Aries.  The Aries cloud in this diagram was plotted
at the wrong distance from Hobbs et al.  (1986), relating it to the
Taurus complex.  In reality, the Aries cloud is so distant from the
Galactic plain, that it hardly can be related both to the Taurus and
to the Perseus dust layers.  Its radial velocity, determined from radio
lines of CO, is also different (Ungerechts \& Thaddeus 1987).
\vskip0.5mm

In a distance range of 250--400 pc, i.e., within the error box of the
cloud distance, the following 11 stars have extinctions $A_V$ larger
than 0.8 mag:  66 (F5 V), 93 (G0 IV-V), 126 (F0 V), 149 (F6 V), 119 (F0
IV), 26 (F6 V), 39 (G5 IV), 116 (M1 III), 139 (G7 III), 62 (G2 IV-V),
and 146 (F4 V).  These stars are candidates to populate either the dust
cloud or the close vicinity behind it.  It is important to find their
precise distances by increasing the accuracy of their absolute
magnitudes.  Also, these stars should be verified carefully for the
presence of secondary components.  In this respect, radial velocity
measurements would be helpful.
\vskip0.5mm

On the other hand, a more exact determination of the cloud distance may
be obtained by increasing the number of investigated stars, i.e., by
shifting the limiting magnitude of the program to fainter stars.  The
program also should include more stars projected onto the dark cloud.
\vskip0.5mm

Additionally, in the present study we have detected about ten stars with
unusual photometric properties.  Due to peculiar energy distribution,
they cannot be classified from their photometric $Q$-parameters.  Short
information about these stars is given in the Notes to Table 1.
Probably, the majority of them are unresolved binaries.  However, in
some cases other causes of their peculiarity may be responsible.  For
example, the star No. 16 seems to be a metallic-line star, No. 69
photometrically is most similar to a R-type carbon star, and No. 95 may
be a F subdwarf.
\vskip0.5mm

Additional information on the amount of dust in the cloud and its
properties can be obtained by polarization studies of the stars.
Andersson (2002) has made polarimetry in the $R$ passband for almost all
stars of Table 1. Smaller number of stars has been measured earlier by
Bhatt \& Jain (1992), Leroy (1993) and Andersson \& Wannier (1995).
The stars of our sample with the largest color excesses exhibit the
largest polarization (up to 3.7\%).  The heavily reddened stars found in
the study may be used in future for more careful investigation of
interstellar reddening and polarization law in the dust cloud.

\vskip5mm

ACKNOWLEDGMENTS.  The investigation is partly supported from the AAS
Chretien Grant of 2000.  We are grateful to the Arizona University
telescope time allocation committee for the observing time on Mount
Lemmon, to the Vatican Observatory community for their help and
hospitality during the observing run, to B.-G.  Andersson (The John
Hopkins University, Baltimore) for the proposal of the present study and
useful discussions, to A. G. Davis Philip (Union College and the
Institute for Space Observations, Schenectady) for reading the
manuscript and for important comments.  We acknowledge the use of the
Simbad database of the Strasbourg Stellar Data Center.

\References

\ref Andersson B.-G. 2002, in preparation

\ref Andersson B.-G., Wannier P. G. 1995, ApJ, 443, L49

\ref Bhatt H. C., Jain S. K. 1992, MNRAS, 257, 57

\ref Bhatt H. C., Sagar R. et al. 1994, A\&A, 289, 946

\ref Cardelli J. A., Clayton G. C., Mathis J. S. 1988, ApJ, 329, L33

\ref Cardelli J. A., Clayton G. C., Mathis J. S. 1989, ApJ, 345, 245





\ref Duerr R., Craine E. R. 1982, AJ, 87, 408

\ref Hearty T., Neuh\"auser R., Stelzer B., Fernandez M., Alcala J. M.,
Covino E., Hambaryan V. 2000a, A\&A, 353, 1044

\ref Hearty T., Fernandez M., Alcala J. M., Covino E., Neuh\"auser R.
2000b, A\&A, 357, 681

\ref Hobbs L. M., Blitz L., Magnani L. 1986, ApJ, 306, L109

\ref Hobbs L. M., Blitz L., Penprase B. E., Magnani L., Welty D. E.
1988, ApJ, 327, 356

\ref Jayawardhana R., Wolk S. J., Navascues D. B., Telesco C. M., Hearty
T. J. 2001, ApJ, 550, L197

\ref Kazlauskas A., \v Cernis K., Laugalys V., Strai\v zys V. 2002,
Baltic \hfil\break Astronomy, 11, 219 (Paper I, this issue)

\ref Leroy J. L. 1993, A\&AS, 101, 551

\ref Luhman K. L. 2001, ApJ, 560, 287

\ref Lynds B. T. 1962, ApJS, 7, 1

\ref Magnani L., Blitz L., Mundy L. 1985, ApJ, 295, 402


\ref Moriarty-Schieven, Andersson B.-G., Wannier P. G. 1997, ApJ, 475,
642


\ref Parenago P. P. 1945, AZh, 22, 129

\ref Perryman M. A. C., Brown A. G. A., Lebreton Y. et al. 1998, A\&A,
331, 81

\ref Pound M. W., Bania T. M., Wilson R. W. 1990, ApJ, 351, 165

\ref Sharov A. S. 1963, AZh, 40, 900 = Soviet Astronomy, 7, No. 5

\ref Skrutskie M. F., Schneider S. E., Stiening R., Strom S. E. et al.
1997, in {\itl The Impact of Large Scale Near-IR Sky Surveys}, eds. F.
Garzon et al., Kluwer Publishing Company, Dordrecht, p. 25

\ref Strai\v zys V. 1977, {\itl Multicolor Stellar Photometry}, Mokslas
Publishers, \hbox{Vilnius,} Lithuania


\ref Strai\v zys V. 1992, {\itl Multicolor Stellar Photometry}, Pachart
Publishing House, Tucson, Arizona

\ref Strai\v zys V., \v Cernis K., Barta\v si\= ut\.e S. 2001a, Baltic
Astronomy, 10, 319

\ref Strai\v zys V., \v Cernis K., Barta\v si\= ut\.e S. 2001b, A\&A,
374, 288

\ref Strai\v zys V., \v Cernis K., Barta\v si\= ut\.e S. 1996, Baltic
Astronomy, 5, 125





\ref Strai\v zys V., Kazlauskas A. 1993, Baltic Astronomy, 2, 1


\ref Strai\v zys V., Kurilien\.e G., Jodinskien\.e E. 1982, Bull.
Vilnius Obs., No. 60, 3


\ref Ungerechts H., Thaddeus P. 1987, ApJS, 63, 645






\ref Zimmerman T., Ungerechts H. 1990, A\&A, 238, 337

\bye

\vskip5mm

Fig. 1. Identification chart for the investigated area. The stars are
numerated according to their RA (2000).

\vskip3mm

Fig. 2. The extinction $A_V$ plotted as a function of the distance $r$
in parsecs.  The solid line shows the extinction run according to the
Parenago formula for the --34$\degr$ Galactic latitude.  The broken
line is the limiting curve for stars of $V_{\rm {lim}}$ = 12 mag and
$M_V$ = +1 mag.  Above this curve a strong selection effect is present.
For more details see the text.  The circle with the error bar is shown
at the estimated distance of the dust cloud.

\bye

%% file: balticw.tex
%
%
\magnification=\magstep1
\baselineskip=11pt plus .1pt minus .1pt
\hsize=12.5truecm
\vsize=19.0truecm  
\hfuzz=5pt\vfuzz=5pt
\tolerance=1000
\overfullrule=0pt
\parskip=0pt
\abovedisplayskip=3 mm plus6pt minus 4pt
\belowdisplayskip=3 mm plus6pt minus 4pt
\abovedisplayshortskip=0mm plus6pt minus 2pt
\belowdisplayshortskip=2 mm plus4pt minus 4pt
\predisplaypenalty=0
\clubpenalty=10000
\widowpenalty=10000
\parindent=2em
%
%
\font\pgnumfont=cmr9
\font\headlinefont=cmti9
 \font\titlefont=cmbx10
\font\authorfont=cmr10
\font\addressfont=cmti9
\font\datefont=cmr9
\font\sumfont=cmr9
\font\itl=cmti9

\font\rml=cmr9

\font\absfont=cmbx9
\font\secfont=cmr10
\font\subsecfont=cmti10
\font\subsubsecfont=cmr10
\font\figfont=cmr9
\font\figheadfont=cmbx9
\font\tabfont=cmr9
\font\tabheadfont=cmbx9
\font\mainfont=cmr10
\font\petitrm=cmr9

%
%
%
\newtoks\TITLE \newtoks\AUTHOR \newtoks\ADDRESS \newtoks\SUMMARY
\newdimen\sumindent \sumindent=\parindent
\newtoks\KEYWORDS \newtoks\SUBMITTED \newtoks\ACCEPTED
\newtoks\SENDOFF
%

%
%
\newtoks\firstpage
\let\firstpage=Y
\newtoks\AUTHORHEAD \newtoks\ARTHEAD \newtoks\VOLUME \newtoks\PAGES
\if!\the\AUTHORHEAD!\AUTHORHEAD={\the\AUTHOR}\fi
\if!\the\ARTHEAD!\ARTHEAD={\the\TITLE}\fi
\footline={\hfil}
\headline={\ifodd\pageno\rightheadline \else\leftheadline\fi}
\def\leftheadline{\if Y\firstpage\firsthead\global\let\firstpage=N
  \else\lefthead\fi}
\def\rightheadline{\if Y\firstpage\firsthead\global\let\firstpage=N
  \else\righthead\fi}
\def\lefthead{\pgnumfont\number\pageno\hfil\headlinefont\the\AUTHORHEAD}
\def\righthead{\headlinefont\the\ARTHEAD\hfil\pgnumfont\number\pageno}
\def\firsthead{\headlinefont Baltic Astronomy,~vol.\the\VOLUME,
\the\PAGES,~\the\year .\hfil}
\voffset=2\baselineskip 
%

\newdimen\oldbaselineskip \oldbaselineskip=\baselineskip
\def\test#1{\newlinechar=`@\if!\the#1! \message{#1 not given@}\fi}%
\def\printheader{
  \parindent=0pt
  \null\vskip1.cm
  \test{\TITLE}
  \vbox{\baselineskip=15pt
    \titlefont\the\TITLE
    }
  \vskip8mm plus8mm
  \test{\AUTHOR}
  \authorfont\the\AUTHOR
  \vskip2mm
  \test{\ADDRESS}
  \addressfont\the\ADDRESS
  \vskip2mm
  \test{\SUBMITTED}
  \line{\datefont Received \the\SUBMITTED
    \if!\the\ACCEPTED!\else, accepted \the\ACCEPTED\fi.\hfill}
  \vskip4mm plus4mm
  \vbox{\leftskip=\sumindent\parindent=0pt
    \parskip=5pt
    \absfont Abstract.
    \test{\SUMMARY}
    \sumfont\the\SUMMARY\par
    \absfont Key words:
    \test{\KEYWORDS}
    \sumfont\the\KEYWORDS\par
    }
  \sumfont
  \if!\the\SENDOFF!\else\footnote{}{Send offprint requests to:
 \the\SENDOFF}\fi
  \parindent=2em
  }
%
%
\newdimen\uppergap \newdimen\lowergap
\uppergap=5mm \lowergap=3mm
\newdimen\secind \newdimen\subsecind \newdimen\subsubsecind
\setbox0=\hbox{\secfont 9. }\secind=\wd0
\setbox0=\hbox{\subsecfont 9.9. }\subsecind=\wd0
\setbox0=\hbox{\subsubsecfont 9.9.9. }\subsubsecind=\wd0
\def\section#1{\goodbreak\par\vskip\uppergap
  \noindent\hangindent\secind\hangafter=1\secfont#1
  \vskip\lowergap\mainfont\par\nobreak}
\def\subsection#1{\goodbreak\par\vskip\uppergap
  \noindent\hangindent\subsecind\hangafter=1\subsecfont#1
  \vskip\lowergap\mainfont\par\nobreak}
\def\subsubsection#1{\goodbreak\par\vskip\uppergap
  \noindent\hangindent\subsubsecind\hangafter=1\subsubsecfont#1
  \vskip\lowergap\mainfont\par\nobreak}
%
%
\def\WFigure#1#2#3{\goodbreak\midinsert\vbox{
  \null\centerline{#2}\vskip5truemm
  \figheadfont\indent Fig.~#1.\figfont\ #3
  \par\mainfont
  }\endinsert}
%

%

%

%

%
\newdimen\tabind
\setbox0=\hbox{\tabheadfont Table 55.} \tabind=\wd0
\def\Table#1#2{\noindent
  \hangindent\tabind\hangafter=1
  \tabheadfont Table~#1.\tabfont #2
 \par
  \mainfont
  }
%
%
\def\References{\vskip\uppergap
\line{\secfont REFERENCES\hfill}
  \vskip0.8\lowergap
 \petitrm
  }
\def\ref{\goodbreak
\hangindent12pt\hangafter=1
\noindent\ignorespaces}
\def\endref{\egroup}
%
%
\def\byebye{\egroup\par\vfill\supereject\end}
%
%

%
%

\def\degr{\hbox{$^\circ$}}

\def\arcmin{\hbox{$^\prime$}}
\def\arcsec{\hbox{$^{\prime\prime}$}}
\def\utw{\smash{\rlap{\lower5pt\hbox{$\sim$}}}}
\def\udtw{\smash{\rlap{\lower6pt\hbox{$\approx$}}}}


\font\tabfont=cmr9



\def\ddown{\lower2.5ex\hbox}
\def\ddow{\lower1.7ex\hbox}
\def\down{\lower1ex\hbox}
\def\uppp{\raise1ex\hbox}
\def\dnnn{\lower1ex\hbox}
\def\uuppp{\raise2ex\hbox}

\def\ts{\thinspace}
\def\(o-c){$O-C$}


\def\angstr{A\kern-.56em\raise1.9ex\hbox{$\scriptscriptstyle\circ$}$\,$}

\newdimen\free\newdimen\shift
\def\Entry#1#2#3{\par\goodbreak\smallskip%
  \setbox1=\vbox{\advance\hsize by-10mm\parindent=0pt
    \def\\{\par}%
    \it#1. \rm#2}
  \line{\box1\hfill#3}\smallskip
}%
\newdimen\savesize

\def\shiftfigure #1#2#3#4#5{
    \vbox to #2 { \ifodd #5 \rightskip#4 \else\leftskip#4 \fi
                  \null\vfil
                  \figheadfont Fig.~#1.\figfont #3
                  \medskip
                }
                          }

\year2002